\input jnl.itp
\input reforder.itp 

\def\wig#1{\mathrel{\hbox{\hbox to 0pt{%
          \lower.5ex\hbox{$\sim$}\hss}\raise.4ex\hbox{$#1$}}}}
\def\msol{M_\odot}
\def\sss{\scriptscriptstyle}

\title MODELING PRESSURE-IONIZATION OF HYDROGEN IN THE CONTEXT OF ASTROPHYSICS

\author {\rm D. SAUMON}
\affil {\it Dept. of Physics \& Astronomy, Vanderbilt University,
            Nashville, TN 37235, USA}

\author{\rm G. CHABRIER}
\affil{Centre de Recherche Astrophysique de Lyon (UMR CNRS 5574),\\
Ecole Normale Sup\'erieure de Lyon, 69364 Lyon Cedex 07, France}

\author{\rm D. J. WAGNER}
\affil{Dept. of Physics, Applied Physics, and Astronomy,\\
            Rensselaer Polytechnic Institute, Troy, NY 12180, USA}

\author{\rm X. XIE}
\affil{Santa Clara Processor Division, Intel Corporation
        2200 Mission College Blvd,  Santa Clara, CA 95052 USA}

\endtopmatter

\noindent
The recent development of techniques for laser-driven shock compression of hydrogen has opened the
door to the experimental determination of its behavior under conditions characteristic of stellar 
and planetary interiors.  The new data probe the equation of state (EOS) of dense hydrogen in the complex
regime of pressure ionization. The structure and evolution of dense astrophysical bodies depend on 
whether the pressure ionization of hydrogen occurs continuously or through a ``plasma phase
transition'' (PPT) between  a molecular state and a plasma state.
For the first time, the new experiments constrain predictions for the PPT.
We show here that the EOS model developed by Saumon 
\& Chabrier can successfully account for the data, and we propose an experiment that should provide 
a definitive test of the predicted PPT of hydrogen. The usefulness of the chemical picture for
computing astrophysical EOS and in modeling pressure ionization is discussed.


\endpage

\vskip 10pt
\noindent
{\bf 1. INTRODUCTION}
\vskip 10pt

\noindent
A series of shock-compression experiments conducted at the Lawrence Livermore National Laboratory 
has revived much interest in the equation of state (EOS) of hydrogen in the regime of pressure ionization. 
Over the last three years, measurement of the shock temperature \refto{Holmes}, the shock
reflectivity \refto{Celliers} and conductivity
\refto{Weir} have been made,  and pressures above 1 Mbar have been achieved repeatedly 
\refto{DaSilva, Collins}.
These experiments directly probe conditions  of pressure and temperature of great astrophysical interest.

Astrophysical objects in which pressure dissociation/ionization of H takes place
range from very-low mass stars ($M \wig< 0.3\,M_\odot$, where $M_\odot$ is
the mass of the Sun) to brown dwarfs (which have masses below $\sim 0.07\,M_\odot$), and extrasolar 
giant planets down to Jupiter ($0.001\,M_\odot$) and Saturn ($0.0003\,M_\odot$). 
All of these objects share the 
same gross composition of $\sim$ 90\% H and $\sim$ 10\% He by atomic fraction.  The thermodynamics 
of hydrogen in the regime of pressure ionization bears on the mechanical and thermal properties of 
these bodies and determines their interior structure and their evolution. It also directly affects
our knowledge of the inner composition of Jupiter and Saturn \refto{GGH} and theories 
of their formation processes.  

The main aspects of the problem of pressure ionization and its astrophysical significance can be grasped by
considering the phase diagram of fluid hydrogen (Figure 1).  At relatively low pressures $(P \wig< 0.1\,$Mbar),
the fluid is fairly ideal.  As the temperature is raised,  molecules dissociate into atoms which then ionize
to form a weakly-coupled plasma.  At pressures above $\sim 1\,$Mbar, non-ideal effects
dominate and hydrogen is pressure ionized into a dense plasma.  The physical conditions in the plasma
are given by the plasma coupling parameter $\Gamma$ and the electron degeneracy parameter $\theta$. 
Figure 1 shows that the pressure-ionized plasma is partially to strongly degenerate ($\theta \wig< 1$) and 
strongly coupled ($\Gamma > 1$).  In the region near $P \sim 1\,$Mbar and $\log T \wig< 5$ the fluid is composed of
a mixture of molecules, atoms, protons and electrons, all strongly interacting.  This  EOS regime is
crossed by the interior profiles of Jupiter, brown dwarfs and very-low-mass stars (dotted lines).  
The relevance of the new shock-compression data is shown by the calculated principal and reflected 
Hugoniots \refto{Note1}
corresponding to the experiments (up to 3.5 Mbar)\refto{Holmes,DaSilva,Collins,Nellis}.

\vskip 10pt
\noindent
{\bf 2. AN EOS BASED ON THE CHEMICAL PICTURE}
\vskip 10pt

\noindent
The work presented here is based on a single, self-consistent free energy model developed a decade ago
which describes a strongly-correlated mixture of H$_2$, H, H$^+$ and electrons \refto{SCPRL,SC1,SC2}. 
This model is based on the so-called ``chemical picture,'' which
assumes that the species considered remain chemically distinct under all conditions.  In practice, this
means that the contributions of the bound states and of the interparticle interactions to the grand partition 
function of the system can be factorized \refto{SCVH}. In reality, interactions and the spectrum of bound states are
coupled and are not strictly factorizable.  When the coupling is weak, 
various corrections to the
exact factorizability can be applied very successfully.  In the regime of pressure ionization, which,
by definition, is characterized by very strong coupling between interactions and bound states, the
chemical picture breaks down.

On the other hand, equation of state models based
on first-principle approaches consider only electrons and nuclei and solve the Schr\"odinger equation for 
bound and free electronic states of
a N-body Coulomb system within some approximations.  Examples of such ``physical picture'' models are the activity
expansion \refto{Rogers} and quantum Monte Carlo simulations \refto{Magro}.
With increased sophistication, the physical picture is expected
to ultimately provide an ``exact'' description of the phenomenon of pressure ionization. 
In low-mass stars, brown dwarfs and giant planets, the range of physical conditions
encountered compellingly points to an EOS based on the chemical picture, however.  These bodies
span a wide range of EOS regimes (see Figure 1) which no single model based on the physical picture can presently
accommodate.  Since the chemical picture is known to work extremely well at relatively low densities
\refto{Rogers2,Dappen}, for the pure molecular fluid \refto{RRY}, and that it reduces 
to the physical picture for the fully ionized plasma,  it is still
the most attractive approach for generating an EOS for many astrophysical applications. For this reason,
nearly all EOS's developed over the past 30 years for applications to stellar interiors have been based 
on the chemical picture.

Since its conception about forty years ago \refto{Harris}, EOS's based on the chemical picture have
become rather sophisticated in their treatment 
of the coupling between  interactions and bound states, resulting
in very accurate EOS's at relatively low densities \refto{MHD}.  Difficulties arise when
these models are pushed into the strongly coupled regimes where approximations valid at low density
become inaccurate, diverge or lead to unphysical results.  An example is the description of
interactions between particles with bound states using pair potentials at low densities while it
is known that many-body effects become important at high densities.
Within the chemical picture, many-body effects can be accounted for by using {\it effective} pair potentials
obtained by fitting experimental data with a particular EOS model.
In this procedure, any physics missing from the EOS model becomes
absorbed in the effective potential.  It follows that a carefully-constructed chemical picture EOS model, 
fitted to 
all the relevant experimental data  (which currently cover a very limited part of the phase diagram),
will provide a reasonably accurate EOS for astrophysical purposes.  Such a model represents a sophisticated
and physically-based form of ``interpolation'' across the regime of pressure-ionization.  
The reliability of the EOS in this regime depends entirely on the availability
of experimental data in appropriate regimes of pressure and temperature.  In this context, the recent work 
on shock-compression of deuterium fills a great void.

\vskip 10pt
\noindent
{\bf 2.1. The Free Energy Model}
\vskip 10pt
\noindent
The free energy model developed by Saumon and Chabrier (SC) and the resulting hydrogen EOS
have been described thoroughly elsewhere \refto{SC1, SC2, SCVH} and only a brief outline will be
given here.  Within the $(P,T)$ region shown in Figure 1, the EOS is derived from 
a single expression for the Helmholtz free energy of a mixture of H$_2$, H, H$^+$ and electrons.
The main features of the model include 1) Finite-temperature Fermi-Dirac statistics for the
electrons; 2) Realistic interaction potentials for the neutral particles (H$_2$ and H), with the
correlation free energy described with a two-component fluid perturbation theory; 3) A detailed treatment
of the internal partition function of H and H$_2$; including 4) An occupation probability
formalism to describe the effects of interparticle interactions on bound states \refto{HM}; 
5) The Two-Component Plasma  model (TCP)
describes the low-density, fully ionized plasma; 6) The high density plasma is described with a Screened
One Component Plasma model (SOCP); in which 7) The electron screening is described by the linear 
perturbation theory using a
finite-temperature dielectric function and a local field correction; 8) Exchange and correlation terms
for the quantum electron fluid; and 9) A polarization potential for the interaction between charged and
neutral particles.  
The chemical equilibrium of the system  at a given density and temperature is obtained by minimizing the 
fully non-ideal Helmholtz free energy.  All equilibrium thermodynamic quantities can then be obtained by
differentiation of the free energy using well-known expressions.

Central to the calculation of the thermodynamics of the dense, partially dissociated/ionized fluid
are the three potentials $\phi_{ij}(r)$
between the neutral particles ($\lbrace i,j \rbrace \equiv \lbrace {\rm H,H}_2\rbrace$).
They can be obtained from {\it ab initio} quantum mechanical calculations for a pair of
particles ({\it e.g.} two molecules).  Such potentials, which do not include the many-body effects,
are known to be too repulsive at short range \refto{RRY}.  At the time the SC model was developed, the 
only available effective potential was $\phi_{\sss {\rm H}_2-{\rm H}_2}$, derived by fitting a thermodynamic 
model to shock-compression data 
\refto{RRY}. {\it Ab initio} potentials were used for $\phi_{\sss {\rm H}_2-{\rm H}}$ and $\phi_{\sss {\rm H}-{\rm H}}$
\refto{PK, KW}. 

\vfill\eject
\noindent
{\bf 2.2. Pressure Ionization}
\vskip 10pt
\noindent
The most remarkable prediction of this EOS model is that pressure ionization of H occurs through
a first-order phase transition between an insulating molecular phase and a partially-ionized, mostly dissociated,
 conducting
phase \refto{SCPRL,SC1,SC2}.  
This so-called plasma phase transition (PPT) is a robust prediction
of the model. No reasonable variation of the model, 
such as changes in the neutral-charged interactions and in the interaction potentials between H and H$_2$,
has produced an EOS without a PPT \refto{SC2},
A qualitatively similar transition is also found in Quantum Monte Carlo 
simulations of hydrogen \refto{Magro}.  

Whether pressure ionization of H occurs continuously or through a PPT
remains one of the major unanswered questions in our understanding of the properties of matter under extreme 
conditions.  The answer has profound astrophysical implications. The PPT, if it exists,
takes place in the interiors of brown dwarfs and giant planets (Figure 1) and would have  important 
consequences on their structure and evolution \refto{SS,CSHL,SHCV}. 
Unfortunately, no present experimental evidence directly addresses this question.  Conductivity measurements of
shock-compressed H$_2$ and D$_2$ up to 1.8 Mbar show a {\it continuous} transition in the
measured electrical conductivity from a semi-conducting to a conducting 
state \refto{Weir}. This result, however, does not preclude the existence of a
{\it structural} transition like the PPT. The increase in conductivity occurs in a
very weakly dissociated phase (dissociation fraction $\sim 5\%$), revealing a
conducting {\it molecular} phase. This is consistent with electronic conduction caused by electrons delocalized
from H$_2^+$ ions, a feature reminiscent of the decreasing band
gap with increasing pressure in solid H$_2$ \refto{MH}.
The conductivity remains constant above 1.4 Mbar up to the largest pressure reached (1.8 Mbar) but is 
significantly
smaller than that expected from a {\it fully dissociated} H plasma. This strongly suggests that
the monoatomic metallic state lies at pressures higher still, a possibility entirely consistent with
our PPT calculation. 

A ``linear-mixing'' model has been formulated by Ross \refto{Holmes, Ross, Zinamon} to explain the general 
features revealed by recent shock-compression experiments which are discussed below.  This model interpolates 
between a purely
molecular model and a fully ionized monoatomic model, which are the proper limits of the data at low
and high densities, respectively.   This interpolation is done by a linear mixing of the molecular and
metallic free energies, including an ideal entropy of mixing term and adjusting an entropy parameter to fit the
shock-compression data \refto{Holmes}.  While the molecular and metallic limits of this model share similarities 
with the
SC free energy, it predicts a continuous transition between the molecular state and the plasma state.
This is because the linear-mixing model excludes any critical behavior {\it by construction}.

\vskip 10pt
\noindent
{\bf 3. COMPARISON WITH SHOCK-COMPRESSION EXPERIMENTS}
\vskip 10pt
\noindent
{\bf 3.1 The Original EOS Model}
\vskip 10pt

\noindent
Figures 2 and 3 show the PPT (dotted line) and compare the calculated principal Hugoniot of deuterium
(dashed line)
\refto{SCPRL, SC1,SC2,SCVH} with the experimental $(P,\rho,T)$ data \refto{Holmes,DaSilva,Collins,Collins2}.  
The calculated
Hugoniot is just subcritical and shows significant density and temperature jumps across the 
PPT.
While the agreement of the theory with 
the $(P,T)$ data is very good, a systematic shift to lower densities is seen for $P > 1.5\,$Mbar in Figure
2.  Ionization increases along the Hugoniot above 1$\,$Mbar and the Hugoniot asymptotically reaches the 
limiting
compression ratio between final and initial states of $\rho/\rho_{\scriptscriptstyle 0}=4$ 
($0.68\,$g/cm$^3$ here), the value for a 
fully-ionized, high temperature plasma.  
The maximum compression ratio of the experimental Hugoniot ($\sim 5.$8) agrees remarkably well with the
prediction of the model.  This generally good agreement validates the treatment of non-ideal effects and pressure
dissociation/ionization in the SC EOS.
Even considering the experimental uncertainties, however, the $(P,\rho)$ data show no evidence of an upturn 
in the Hugoniot at $\rho \approx 0.7\,$g/cm$^3$ or for the discontinuity across the PPT (Figure 2).  
This indicates that
the critical point must be below the calculated value of $T_c=15300\,$K, and $P_c=0.615\,$Mbar 
\refto{SC2}.
These experimental results are the very first to provide a constraint on any calculation of the PPT.

\noindent
{\bf 3.2 Effective Pair Potentials Fitted to Temperature Data}
\vskip 10pt

\noindent
The first measurements of the temperature of shock-compressed deuterium are compared to the single-
and double-shock Hugoniots computed from the original SC EOS \refto{SCVH} in Figure 4 (the same Hugoniots
are also shown in Figure 1).  The reflected-shock 
temperatures \refto{Holmes} are about 30\% below the values predicted by the original model EOS (dotted curve).  
The same Hugoniots are also shown in Figure 1.  These measurements clearly indicate that
the kinetic energy of the shock is partly absorbed in the dissociation of molecules rather than being
transferred to their thermal motion. 
In this experiment \refto{Holmes}, the dissociation fraction reaches 20--25\%; large enough to
allow a determination of {\it effective} H$_2$--H  and H--H potentials.  Thus, considerably softer 
effective pair 
potentials $\phi_{\sss {\rm H}_2-{\rm H}}(r)$ and $\phi_{\sss {\rm H}-{\rm H}}(r)$ were obtained by fitting 
the temperature data shown in Figure 4.  The resulting Hugoniots are shown by solid lines
in Figure 4. The figure shows three double-shock Hugoniots and their corresponding initial states along
the single-shock Hugoniot (three of the data points repeat the same measurement). The new effective
potentials give a very good agreement with the data, except for the double-shock point represented by a triangle,
which may be anomalous (a similarly poor fit of this point is found in the original analysis of this 
data \refto{Holmes}).

The PPT coexistence curve and principal Hugoniot calculated with the same Helmholtz free energy model but 
using the new effective potentials are shown as solid lines in Figures 2 and 3.  The PPT is mildly affected
and the critical point is now at $T_c= 14600\,$K and $P_c=0.730\,$Mbar.  The new Hugoniot is
barely supercritical and is therefore continuous.  In the $(P,T)$ plane, the Hugoniot
remains in excellent agreement with the data and  the agreement with the $(P,\rho)$ data is much improved.
We emphasize that the Hugoniot shown by the solid curves in Figures 2 and 3 is not  fitted to the
data shown in these figures, but obtained by fitting the 
$\phi_{\sss {\rm H}_2-{\rm H}}(r)$ and $\phi_{\sss {\rm H}-{\rm H}}(r)$ to an independent data
set \refto{Holmes}  measured in a different $(P,T,\rho)$ regime (see Figures 1 and 4).  The
linear-mixing model of Ross \refto{Ross}, which is also adjusted to fit the temperature measurements
shown in Figure 4, reproduces the experimental single-shock Hugoniot of Figures 2 and 3 just as well as our
model \refto{Collins, Collins2}.

For the first time, experimental data which probe the $(P,T)$ 
domain characteristic of the PPT are available.  They clearly show that the
earlier determination of the critical point \refto{SCPRL, SC1, SC2}, based on potentials determined from 
experiments done 16 years ago
\refto{Nellis,RRY} overestimated $T_c$.   Figure 2 suggests that the revised critical point may still be
somewhat too hot.  The ability to make such statements represents a major step forward in our
understanding of the pressure ionization of hydrogen.

The new techniques developed to reach pressures of several Mbar in shock compression experiments \refto{DaSilva}
open the possibility of a direct detection of the PPT by generating a reflected shock \refto{Collinspriv}.
A shock reflected from the principal Hugoniot at $P \sim 0.36\,$Mbar and reaching
pressures of up to 3$\,$Mbar would provide a conclusive test of the existence of the PPT as predicted 
by this model.
The reflected Hugoniot reaches the PPT at $T=9000\,$K, well below the calculated critical point.
The density and temperature discontinuities in the reflected Hugoniot at the transition are
$0.6\,$g/cm$^3$ and 1700$\,$K, respectively, which should be larger than the experimental uncertainties.

\vskip 10pt
\noindent
{\bf 4. FUTURE DEVELOPMENTS}
\vskip 10pt

\noindent
The most significant improvement envisioned for the model is the introduction of the dissolution of the upper
bound states of the hydrogen atom by interactions with the surrounding plasma.  This can be done
following a formalism describing the Stark ionization of these states by the fluctuating microfield \refto{SCVH,HM}.
This physical mechanism is important in the regime of temperature ionization
even at low densities and will bring the SC EOS in excellent agreement with the very accurate activity expansion EOS
\refto{Rogers}.

The most natural way to improve the accuracy of this EOS in the regime of pressure ionization is to further
modify the H--H interaction potential.  Similarly, the coupling between charged and neutral
particles (a polarization potential here) could be modified,  although several alternatives have already
been explored \refto{SC2}.  This should allow a better fit of the data shown in Figures 2 and 3.
For the reasons outlined in \S 2, there is little to gain by increasing the level of sophistication of the
model in an attempt to better model pressure ionization within the chemical picture.

The recent experiments \refto{Collins, Collins2}  have revealed that  the current
theories for dense fluid hydrogen strongly  disagree with each other in the regime of pressure
dissociation/ionization.  With the exception of
this work and of the linear-mixing model \refto{Ross}, they agree poorly with the 
data. The guidance of experimental data is therefore  essential to the development of a good 
understanding of this phenomenon.  
This need is most acute for models based on the chemical picture, which depend on experiments for the
determination of effective pair potentials.  The chemical picture is well-suited for the computation
of large EOS tables for astrophysical applications, but models must be calibrated with experimental 
data if they are to be reliable in the difficult regime of pressure ionization.  From this perspective,
the recent shock-compression data are extremely valuable and it is hoped that 
this part of the phase diagram will soon be better charted by new experiments.

{\bf\it Acknowledgments:} We are grateful to R. Cauble and G. Collins for kindly providing their data
in advance of publication and to F. Rogers and W. Nellis for useful discussions.  This 
work was supported in part by NSF grant AST93-18970.

\noindent

\endpage

\references
\refis{Celliers} Celliers, P. M., Collins, G.W., Da Silva, L.B., Gold, D. M., Cauble, R., Wallace, R. J.
                and Foord, M. E. (1999). submitted to {\it Phys. Rev. Letters}.

\refis{CSHL} Chabrier, G., Saumon, D., Hubbard, W. B. and  Lunine, J. I. (1992), {\it  Astrophys. J.}, {\bf 391}, 817.

\refis{Collins2} Collins, G. W. {\it et al.} (1999). submitted to {\it Phys. Rev. Lett.}
                 
\refis{Collins} Collins, G. W. {\it et al.} (1998).  {\it Science}, {\bf 281}, 1178.

\refis{Dappen} D\"appen, W. (1994). {\it Helioseismology: the Sun as a Strongly-Constrained, Weakly-Coupled Plasma},
               p.368, In: {\it The Equation of State in Astrophysics}, Eds. Chabrier, G. and Schatzman, E., 
               Cambridge, Great Britain.

\refis{DaSilva} Da Silva L. B. {\it et al.} (1997).  {\it Phys. Rev. Lett.}, {\bf 78}, 483.

\refis{GGH} Guillot, T., Gautier, D. and Hubbard, W. B. (1997),  {\it Icarus}, {\bf 130}, 534.

\refis{Harris} Harris, G. M. (1959). {\it J. Chem. Phys.}, {\bf 31}, 1211.

\refis{Holmes} Holmes, N. C., Ross, M. and Nellis, W. J. (1995). {\it Phys. Rev. B}, {\bf 52}, 15835.

\refis{HM} Hummer, D. G., and Mihalas, D. (1988). {\it Astrophys. J.}, {\bf 331}, 794.

\refis{KW} Kolos, W. and Wolniewicz, L. (1965). {\it J. Chem. Phys.}, {\bf 43}, 2429.

\refis{Magro} Magro, W. R., Ceperley, D. M., Pierleoni, C. and Bernu, B. (1996). {\it Phys. Rev. Lett.}. {\bf 76}, 1240.

\refis{MH} Mao, H. K. and Hemley, R. J. (1994)  {\it Rev. Mod. Phys.}, {\bf 66}, 671.

\refis{MHD} Mihalas, D., D\"appen, W. and Hummer, D. G. (1988). {\it Astrophys. J.}, {\bf 331}, 815.

\refis{Nellis} Nellis, W. J., Mitchell, A. C., van Thiel, M., Devine, G. J. and  Trainor, R. J. (1983). {\it J. Chem.
               Phys.}, {\bf 79}, 1480.

\refis{PK} Porter, R. N. and Karplus, M. (1964). {\it J. Chem. Phys.}, {\bf 40}, 1105.

\refis{Rogers} Rogers, F. J. and Young, D. A. (1997). {\it Phys. Rev. E}, {\bf  56}, 5876.

\refis{Rogers2} Rogers, F. J. (1994). {\it Equation of State of Stellar Plasmas}, p.16, In: {\it The Equation of 
                State in Astrophysics}, Eds. Chabrier, G. and Schatzman, E., Cambridge, Great Britain.

\refis{Ross} Ross, M. (1998). {\it Phys. Rev. B}, {\bf 58}, 669.

\refis{RRY} Ross, M.,  Ree, F. H. and Young, D. A. (1983). {\it J. Chem. Phys.}, {\bf 79}, 1487.

\refis{SCPRL} Saumon, D. and Chabrier, G. (1989). {\it Phys. Rev. Lett.}, {\bf 62}, 2397. 

\refis{SC1} Saumon, D. and Chabrier, G. (1991). {\it Phys. Rev. A}, {\bf 44}, 5122. 

\refis{SC2} Saumon, D. and Chabrier, G. (1992). {\it Phys. Rev. A} {\bf 46}, 2084.

\refis{SS} Stevenson D. J. and Salpeter, E. E. (1977). {\it Astrophys. J. Suppl.}, {\bf 35}, 221.

\refis{SCVH} Saumon, D. and Chabrier, G. and  Van Horn, H. M. (1995). {\it Astrophys. J. Suppl.} {\bf 99}, 713.

\refis{SHCV} Saumon, D., Hubbard, W. B., Chabrier, G. and Van Horn, H. M. (1992).  {\it Astrophys. J.}, {\bf 391}, 827.

\refis{Weir} Weir, S. T.,  Mitchell, A. C. and Nellis, W. J. (1996). {\it Phys. Rev. Lett.}, {\bf 76}, 1860.

\refis{Zinamon} Zinamon, Z.  and Rosenfeld, Y. (1998). {\it Phys. Rev. Lett.}, {\bf 81}, 4668.

\refis{Collinspriv} G.W. Collins (private communication).

\refis{Note1} A Hugoniot is the curve in $(P,\rho,T)$ space which represents the locus of shocked states of the
           material, starting from one initial pre-shock state.

\endreferences

\endpage

\centerline{\bf Figure Captions}

{\bf Figure 1}: Phase diagram of hydrogen.  The coexistence curve of the plasma phase transition
(PPT) appears at the left of center as a heavy solid line which ends at the critical point.  
Curves of constant plasma coupling parameter ($\Gamma$) and electron degeneracy parameter 
($\theta$) are shown.  
The plasma coupling parameter is a measure of the
strength of the Coulomb interaction. It is defined by $\Gamma=e^2/akT$, the ratio of the electrostatic
energy of two protons at the average interionic distance $a$ to their kinetic energy.
The electron degeneracy parameter is the ratio of the temperature to the Fermi temperature, $\theta=T/T_F$.
Regions dominated by molecules, atoms and ionized H are labeled  and delimited by a dashed curve 
which corresponds to 50\% dissociation or ionization.  
The calculated single- and double-shock Hugoniots corresponding to the experiments are show by the 
thin solid curves.  Finally, the thin black dotted lines show the
internal structure profiles of several astrophysical bodies (from left
to right): Jupiter, a $7 \times 10^9\,$yr old brown dwarf of $0.055\,M_\odot$, and a $0.1\,\msol$ star.
The EOS model is invalid in the hashed region.

\bigskip

{\bf Figure 2} : Pressure-density diagram for deuterium showing the single-shock data \refto{Collins},
calculated Hugoniots and the PPT.  The Hugoniot based on the EOS of 
Ref.~\refto{SCVH} and the corresponding metastable region of the
PPT are shown by dashed and dotted curves, respectively.  This Hugoniot is subcritical.
After fitting potentials to the measurements of \refto{Holmes} the new EOS model predicts that the Hugoniot 
is supercritical, as shown by the solid curve. The hashed area shows the new metastable region which
has a higher critical density.  Note that this is a projection of a $(P,T,\rho)$ surface.  With increasing
pressure, the Hugoniots come out of the page while the PPT goes into the page.
\bigskip

{\bf Figure 3} : Same as Fig.~2 but in the temperature-pressure plane.  The coexistence curves of Ref. 
\refto{SCVH} and the new calculation presented here are shown by the heavy dotted and solid lines,
respectively.  The bend in the supercritical
Hugoniot is caused by the negative sign of $dP/dT|_\rho$ in the vicinity of the critical point.  This
behavior is found in some materials and is not thermodynamically forbidden.
\bigskip

{\bf Figure 4}: Fit to the temperature measurements along the single- and double-shock Hugoniots of D$_2$
\refto{Holmes}. Each double-shock point corresponds to one of the first-shock points shown.  Matching pairs
of single and double-shock points are shown with a common symbol.  The specific volume $V$ is not measured
in the experiment and the error bars in $V$ are taken from $(P,V)$ measurements in a
similar experiment \refto{Nellis} and are approximate.  Calculated double shocks reflected from $V=7.98$,
7.44 and 7.10$\,$cm$^3$/mole along the principal Hugoniot are shown.

\endpaper
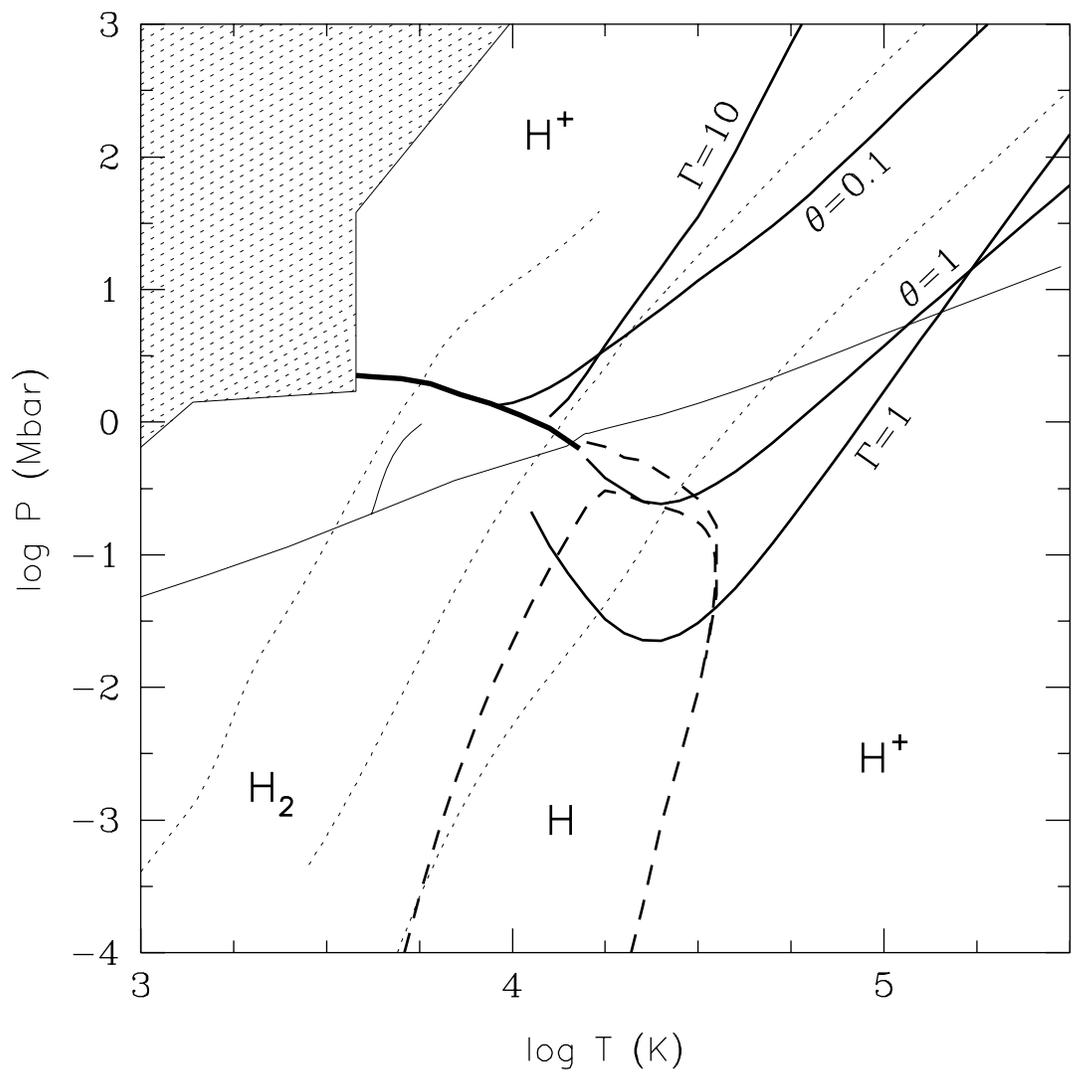
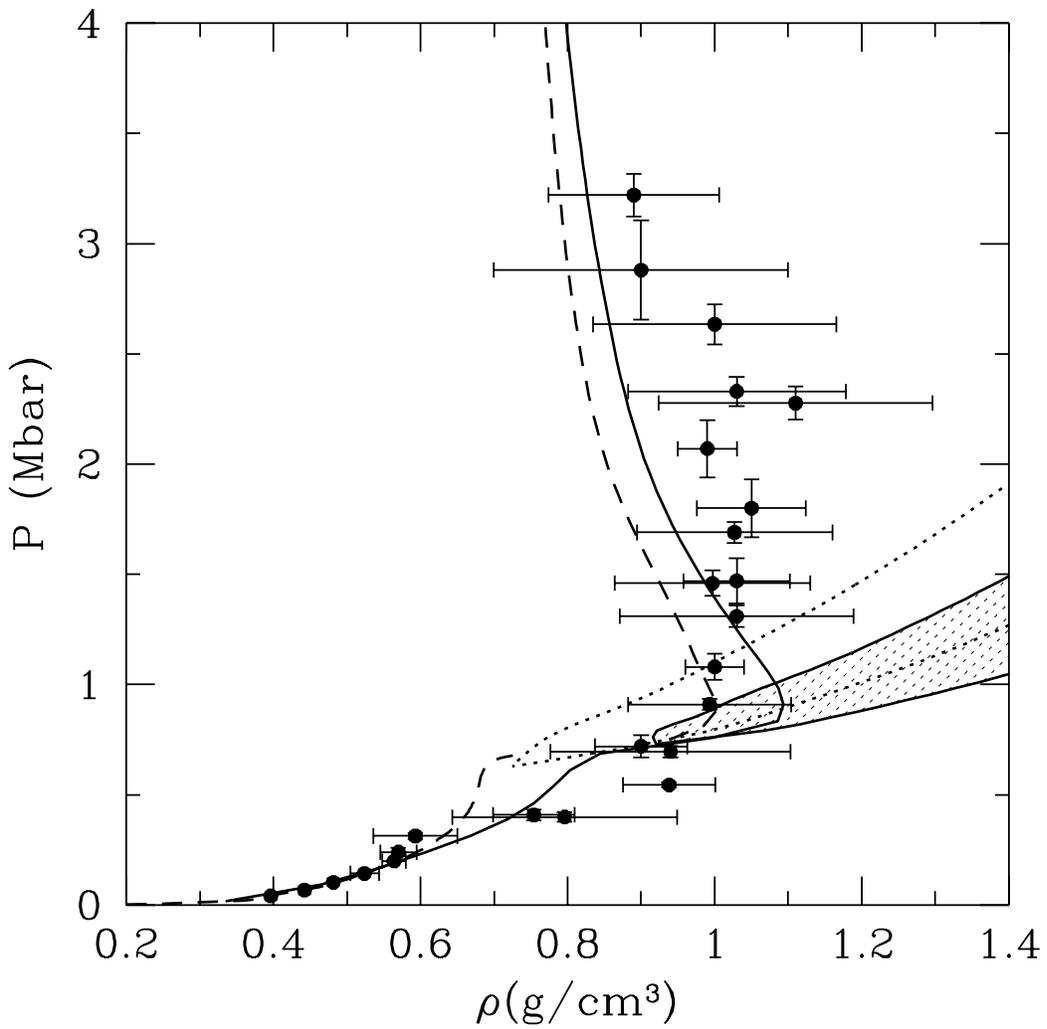
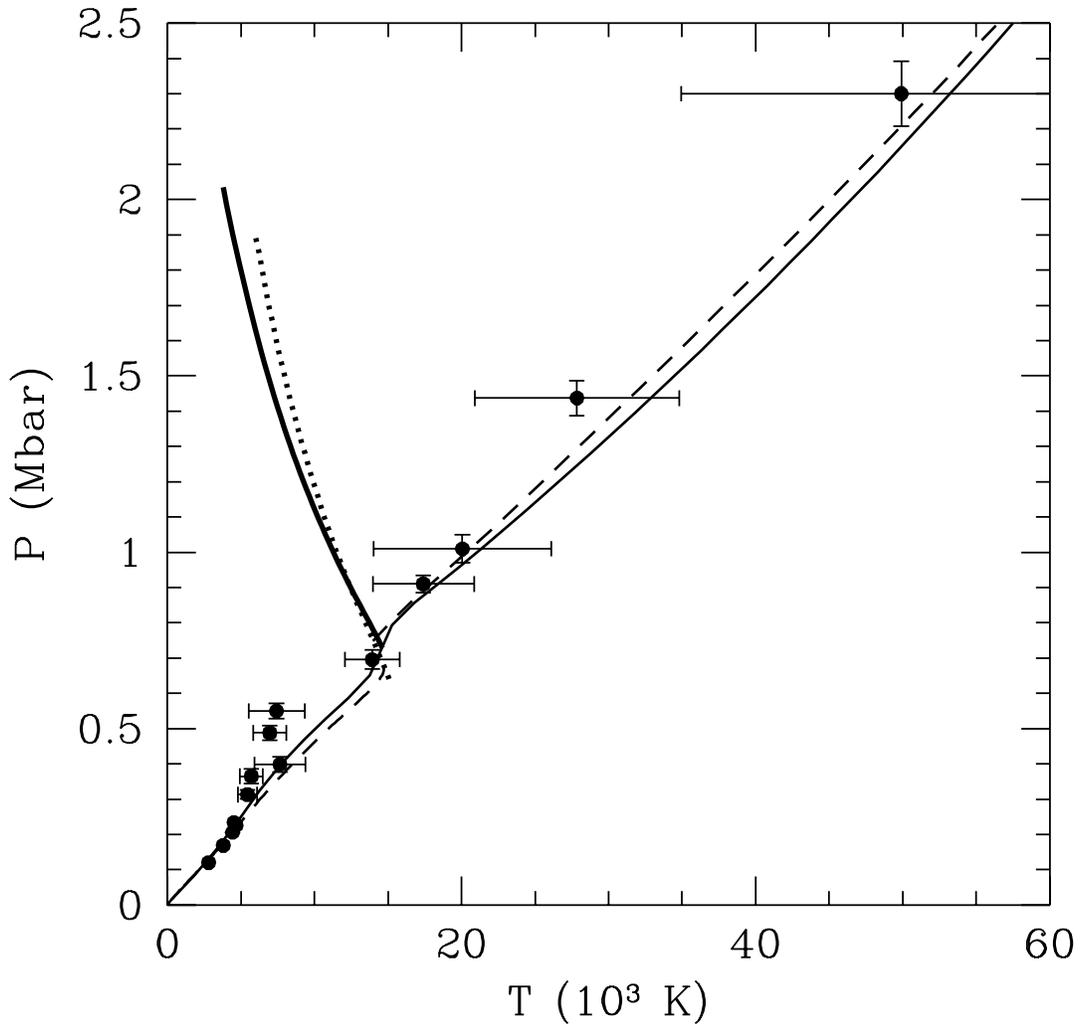
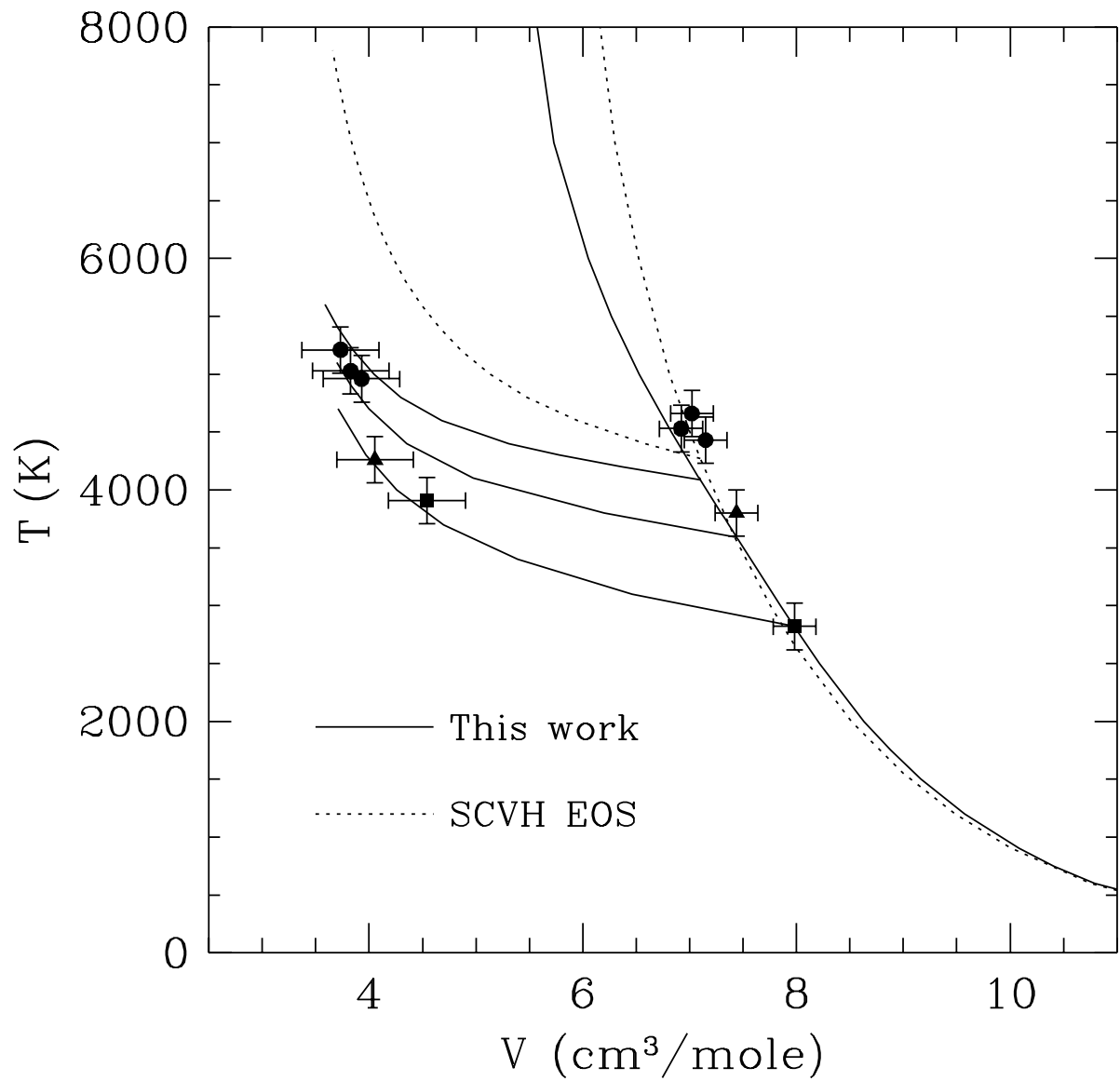
\bye